\documentstyle[preprint,prl,aps,epsf]{revtex}
\tightenlines
\begin{document}
\draft
\title{Quadrupole Deformation of Barium Isotopes}
\author{Michiaki Sugita}
\address{Japan Atomic Energy Research Institute,Tokai, Ibaraki 
319-1195, Japan}
\author{Koji Uchiyama}
\address{
The Institute of Physical and Chemical Research (RIKEN), Wako, Saitama
351-0198, Japan\\
and\\
Institute of Physics, University of Tsukuba, Ibaraki 305-0006, Japan
}
\author{Kohei Furuno}
\address{
  Institute of Physics and Tandem Accelerator Center, University of 
Tsukuba,
  Ibaraki 305-8577, Japan
}
\maketitle

\begin{abstract}
The B(E2:$0_1^+\rightarrow 2_1^+$) values of the Ba
isotopes ($Z=56$) exhibit a sharp increase in deformation as the
neutron numbers approach the mid-shell value of $N=66$.
This behavior
is anomalous because the $2^+_1$ level energies are very similar to
those of the neighboring isotopes.
By means of the axially-symmetric deformed Woods-Saxon (WS) hamiltonian 
plus
the BCS method, we investigated the systematics of B(E2) of the Ba 
isotopes.
We showed that 15\% of the B(E2) values at $N=66$ was due to the
level crossing, occurring at the deformation $\beta_{\hbox{\scriptsize 
WS}}$ $\sim$ 0.3,
between the proton orbits originating
from the orbits $\Omega^\pi=1/2^-(h11/2)$ and $9/2^+(g9/2)$ at zero
deformation.
The latter of these two was an intruder orbit originating from below the
energy gap at $Z=50$, rising higher in energy with the
deformation and intruding the $Z=50-82$ shell.
These two orbits have the largest magnitude of the quadrupole moment 
with
a different sign among the orbits near and below the Fermi surface.
Occupancy and non-occupancy of these orbits by protons thus affect
B(E2:$0_1^+\rightarrow 2_1^+$) significantly.
\end{abstract}

\pacs{PACS Numbers: 21.10.-k; 21.10.Ky; 21.60.Ev; 21.60.Fw; 23.20.-g}

\noindent {Keywords: B(E2), Ba, Woods-Saxon hamiltonian, BCS, IBM-2}

\section{Introduction}
The deformation of the nuclear ground state is a fundamental quantity
that can be deduced from the B(E2) value for the first $2^+$ state
($J^\pi=2_1^+$). From the systematics of the B(E2:$0_1^+\rightarrow 
2_1^+$) values, we could
infer how the deformation of the ground state changes from nucleus to
nucleus.  Over the past decade, the B(E2) values of the Xe-Ba-Ce
nuclei have been measured extensively.  See \cite{Xe} for Xe, \cite{Ba}
for Ba, \cite{Ce} for Ce.  The accumulated data has enabled us to
compare the data with several theoretical results to test these
theories \cite{raman95}.

Focusing ourselves on the Ba (Z=56) isotopes, we will discuss
the systematics of the B(E2:$0_1^+\rightarrow 2_1^+$) values.

\section{Grodzins rule}
The B(E2) values of Ba \cite{uchiyama} as well as the Xe (Z=54) isotopes
\cite{raman95} increase sharply as the neutron number $N$
approaches $N=66$ \cite{raman95,uchiyama} as can be seen in 
Fig.~\ref{fig:Ba}.
The solid squares in Fig.~\ref{fig:Ba} present the experimental data of 
B(E2:$0_1^+\rightarrow 2_1^+$),
which are compared in Fig.~\ref{fig:Ba} with the theoretical results
estimated using the Grodzins
rule \cite{Grodzins} and obtained using the proton-neutron interacting 
boson model
(IBM-2) \cite{pauli}, as we will explain in detail below.

The Grodzins rule \cite{Grodzins} expresses an empirical relation
between Ex($2_1^+$) and B(E2):
\begin{equation}
B(E2:0^+\rightarrow 2^+) =1.63 \times 10^{-2} Z^2 / (E\cdot A) (eb)^2
\label{eq_grodzins}
\end{equation}
with $E$ being Ex($2_1^+$) (MeV), $Z$ the proton number, and $A$ the 
mass number.
In deriving Eq.~(\ref{eq_grodzins}), we used Eq.~(4) of 
Ref.~\cite{Grodzins}.
The open triangles in Fig.~\ref{fig:Ba} present B(E2:$0_1^+\rightarrow 
2_1^+$) derived from Eq.~(\ref{eq_grodzins}).
The results obtained by this rule are in surprisingly good
agreement with the data for the less deformed Ba isotopes with $N\ge 
70$ but
underestimate B(E2) by 20\% at $N=66$.
The similar empirical relation in \cite{raman95} also underestimates
B(E2:$0_1^+\rightarrow 2_1^+$) at $N=66$.
The empirical rules suggest that the sharp increase of B(E2)
is anomalous because Ex($2_1^+$) at $N=66$ is almost the same as those 
of the
neighboring isotopes.
We will next examine this anomaly in B(E2:$0_1^+\rightarrow 2_1^+$) 
within the framework of
IBM \cite{IA}.

\section{IBM-2}
The interacting boson model (IBM) \cite{IA} has been widely used for
describing the quadrupole collective states of the medium heavy nuclei.
The microscopic foundation of IBM has been given within the 
proton-neutron
IBM (IBM-2) \cite{ibm2}.
The building blocks of IBM-2 are $s_\tau$ and $d_\tau$ bosons
($\tau=\pi, \nu$) which are considered to be approximations to the
proton (neutron) pairs with spin-parity $0^+$ and $2^+$.
The boson images of the fermion operators are given in terms of
the OAI mapping  \cite{OAI}.

\begin{figure}
  \epsfxsize=10cm
  \centerline{\epsfbox{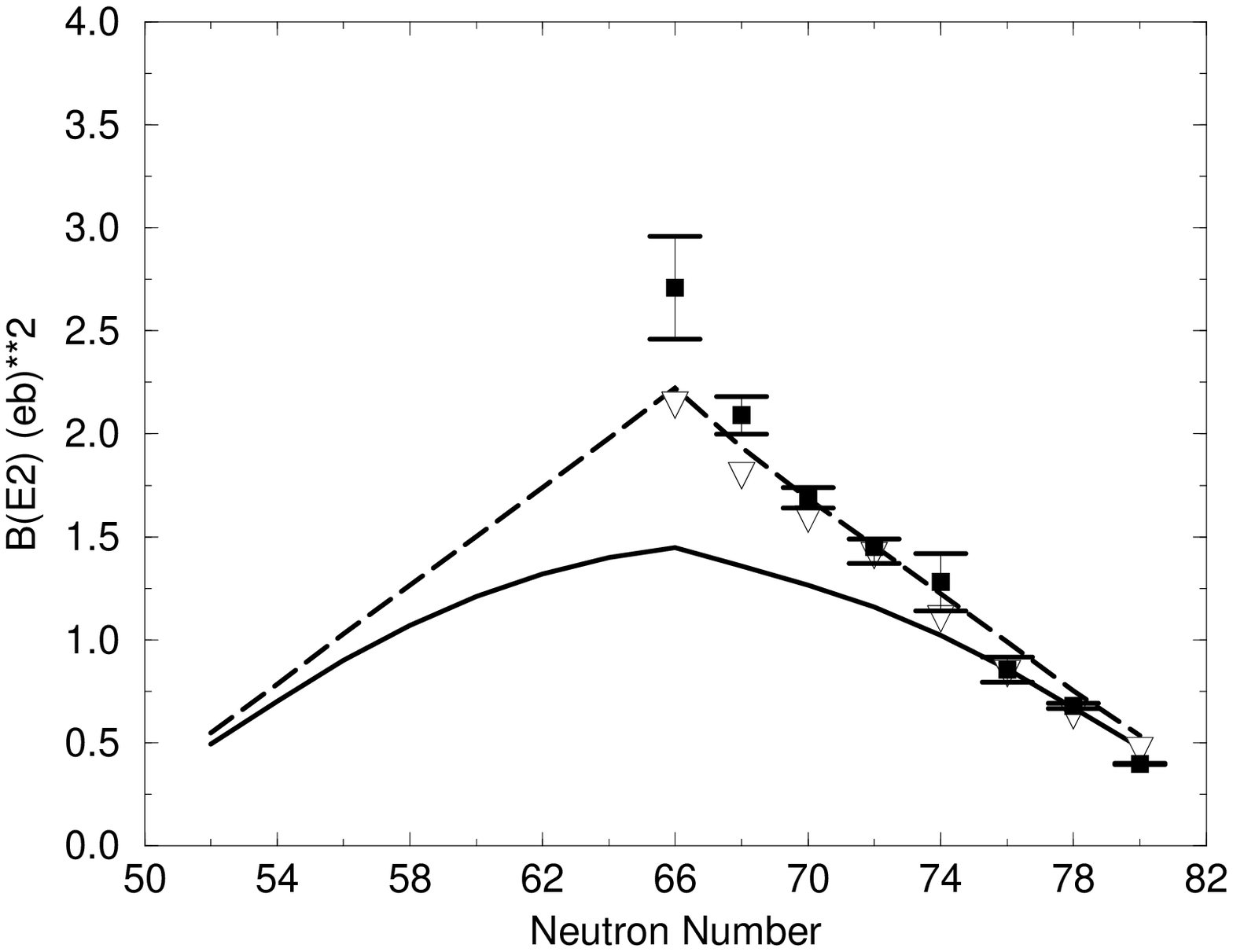}}
  \caption{Systematics of the B(E2:$0_1^+\rightarrow 2_1^+$) values of 
the Ba isotopes.
    The solid squares
    express the experimental data.
    The open triangles present the predictions by the Grodzins rule.
    The solid (dashed) line denotes the IBM-2 results
    with (without) the Pauli factor.}

\label{fig:Ba}
\end{figure}

Using IBM-2, Otsuka, Pan and Arima \cite{pauli} investigated the
Xe-Ba-Ce isotopes concentrating on the systematics of 
B(E2:$0_1^+\rightarrow 2_1^+$) as a
function of the neutron number.  According to the procedures given in
\cite{pauli}, we did the IBM-2 calculations using the code NPBOS
\cite{npbos} with the parameters taken from Ref.\cite{pauli} which were
derived from the microscopic calculations through the method of OAI
mapping \cite{OAI}.  The boson image of the fermion E2 operator is
given by \cite{pauli}
\begin{equation} T^{(E2)}=e_\pi^0 Q_\pi + e_\nu^0 Q_\nu,
\label{eq_e2}
\end{equation}
where $e_\sigma^0$ ($\sigma=\pi, \nu$) are constants and
\begin{equation}
Q_\sigma = \kappa_\sigma \left\{
  d_\sigma^{\dagger}\tilde{s}_\sigma +
  s_\sigma^{\dagger}\tilde{d}_\sigma + \chi_\sigma
  \left[d_\sigma^{\dagger}\tilde{d}_\sigma\right]^{(2)}\right\}
\label{eq_Q}
\end{equation}
with
\begin{equation}
\kappa_\sigma=\sqrt{\frac{(\Omega_\sigma-N_\sigma)}{(\Omega_\sigma-1)}},
\label{eq_kappa}
\end{equation}
\begin{equation}
\chi_\sigma = \chi_\sigma^0
\frac{\Omega_\sigma-2N_\sigma}{\Omega_\sigma-2} \cdot
\kappa_\sigma^{-1}
\label{eq_chi}
\end{equation}
where $\Omega_\sigma$ is the
degeneracy of the major shell and $\Omega_{\pi,\nu}=16$ for the Ba
isotopes.  We used the following parameters taken from \cite{pauli}:
$\chi_\sigma^0=\pm0.672$ ($-$ for the particles and $+$ for the
holes), $e_\pi^0=0.154$ (eb), $e_\nu^0=0.110$ (eb).  Using this E2
operator, we calculated B(E2:$0_1^+\rightarrow 2_1^+$) for the Ba 
isotopes, and compared
them with the experimental data \cite{Ba} in Fig.~\ref{fig:Ba}. The 
solid line
represents the theoretical results which were identical to the results
in \cite{pauli}.

Quantities $\kappa_\sigma$ and $\chi_\sigma/\chi_\sigma^0$ represent
the effects of the Pauli principle, and are referred to as the Pauli
factors in \cite{pauli}. These Pauli factors become unity when there is
no Pauli effect, i.e., $\Omega_\sigma\rightarrow \infty$ \cite{pauli}.
The dashed line in Fig.~\ref{fig:Ba} is the B(E2) values calculated 
without the
Pauli factor $\kappa_\sigma$ in $T^{(E2)}$ \cite{pauli}, i.e., with
$\kappa_\sigma=1$ and $\chi_\sigma = \chi_\sigma^0
(\Omega_\sigma-2N_\sigma)/(\Omega_\sigma-2)$.

In Fig.~\ref{fig:Ba}, we used the latest experimental B(E2) data, some 
of which
\cite{uchiyama} were measured after the publication of \cite{pauli}.  We
should mention again that, besides the experimental data, the IBM-2
results in Fig.~\ref{fig:Ba} are identical to Fig.~4 of \cite{pauli} 
because we
have taken the same parameters as was used in \cite{pauli}.

It is clearly seen from Fig.~\ref{fig:Ba} that the IBM-2 predictions 
with the
Pauli factor are in good agreement with the data for $N\ge 76$.
However, for $70 \le N\le 74$, IBM-2 without the
Pauli factor gives B(E2) values closer to the data than IBM-2
with the Pauli factor.
For $N=66, 68$, the B(E2) values of IBM-2 with
the Pauli factor saturate in the mid-shell
in disagreement with the experimental trends.
At $N=66$, the IBM-2 calculations without the Pauli factor
yield nearly the same magnitude of B(E2:$0_1^+\rightarrow 2_1^+$) as 
the Grodzins rule.
Thus, at $N=66$, not only the Grodzins
rule but also IBM-2 underestimate the B(E2) values.

The Xe isotopes also showed a sharp increase in the B(E2) values at
the mid-shell \cite{Xe}. Raman et al. pointed out in
Ref.~\cite{raman95} that the single-shell model such as the
single-shell Nilsson model failed in reproducing the B(E2) values of
the Xe isotopes at $N\sim$ 64-66, whereas several multi-shell models
correctly predicted these values. They showed in \cite{raman95} that
the rapid rise of the $\Omega=9/2$ (g9/2) oblate orbit for protons as
a function of deformation and its intrusion into the 50-82 shell
suggests a mechanism for increasing the B(E2:$0_1^+\rightarrow 2_1^+$) 
value as a result of
partial emptying of this orbit due to pairing.  However, the detailed
manner by which the mid-shell Xenon isotopes acquire significant
deformation is still unclear, as pointed out in \cite{raman95}.

We aim in this paper at clarifying the origin of the sharp increase of
B(E2:$0_1^+\rightarrow 2_1^+$) observed in the mid-shell Ba isotopes.  
To achieve this
objective, we used the axially symmetric deformed Woods-Saxon (WS) 
hamiltonian
plus the BCS method.  All orbits with eigenenergies lower than the
barrier height \cite{WS} were adopted in the calculations.  Therefore,
we can get rid of the drawbacks \cite{raman95} of the single-shell
models such as the single-shell Nilsson model.  In contrast to the
several theoretical tools examined in \cite{raman95}, we do not
minimize a total hamiltonian to fix the intrinsic state of the ground 
band.
Instead, as a function of deformation parameter 
$\beta_{\hbox{\scriptsize WS}}$ of the WS
potential, we calculated intrinsic quadrupole moment $Q_0$. Then, we
determined $\beta_{\hbox{\scriptsize WS}}$ directly from the 
experimental $Q_0$ values, by
which we could discuss the variation of the intrinsic configuration of
the ground band with the deformation. We will show that 15\% of
B(E2:$0_1^+\rightarrow 2_1^+$) for the Ba isotope with $N=66$ is due to 
the level crossing
between the $\Omega=9/2 (g9/2)$ and $\Omega=1/2 (h11/2)$ orbits, but
the effect of the level crossing on B(E2) fades out rapidly as $N$
goes away from the mid-shell value of $N=66$. This suggests that the
sharp increase in B(E2) at the mid-shell is due to the level crossing
between these orbits.

\section{Woods-Saxon Hamiltonian plus BCS Calculations}
We assume in this paper that individual nucleons move in the
axially-symmetric quadrupole deformed Woods-Saxon potential and
interact with each other by the monopole pairing force. We take the
same form of the Woods-Saxon (WS) potential as was used in \cite{WS}.
We assume $Y_{2,0}$ deformation only, and denote the deformation
parameter of the WS potential by $\beta_{\hbox{\scriptsize WS}}$ as was 
already mentioned.

We should mention here the $\gamma$-softness or triaxiality of the
heavier Ba isotopes with $N<82$. The Ba isotopes exhibit the
$\gamma$-soft or O(6)-like level scheme for the heavier isotopes with
$N<82$ \cite{o6,Puddu}, and change gradually from the
$\gamma$-soft to axially symmetric nuclei as $N$ approaches the
mid-shell value of $N=66$ \cite{Puddu}.  Through this paper, we
assumed the axially symmetric WS potential for simplicity.  This
assumption may not be so valid for the heavier Ba isotopes because of
the $\gamma$-softness. However, our major interest lies in the rapid
increase in B(E2) of the Ba isotopes with $N\sim 66$. For these
nuclei, the axial symmetry is a reasonable approximation.

We use the WSBETA code \cite{WS} for calculating the eigenenergies of 
the
WS hamiltonian as well as the quadrupole moments of the eigenstates.
We have several parameters in the WS potential \cite{WS} such as the
depth, radius and diffuseness of the central potential and
those of the spin-orbit potential. Among several parameter sets of the
WS potential \cite{WS}, we chose the ``universal'' parameter set
\cite{WS}.  The other WS parameter sets such as the Blomqvist-Wahlborn,
Chepurnov, Rost, and ``optimal'' parameter sets \cite{WS} give almost
the same results as the ``universal'' parameter set, although the
details of these
results are not shown in this paper.  Thus, the major results of
this paper on the Ba isotopes do not change
among those WS parameter sets.

We assume the pairing force for the residual interaction.
As our concern is the proton quadrupole moment,
we consider only proton system hereafter.
The pair potential acting on the individual nucleons takes up the form
of
\begin{equation}
V_{\rm pair} = - \Delta \sum_{\nu > 0 }
( a^{\dagger}(\bar{\nu})a^{\dagger}(\nu) +
a(\nu)a(\bar{\nu}) )
\end{equation}
where $a^{\dagger}(\nu)$ ($a(\nu)$) is the creation (annihilation)
operator of a proton in the eigenstate $\nu$ of the WS hamiltonian,
$a^{\dagger}(\bar{\nu})$ ($a(\bar{\nu})$) is the corresponding
time-reversed operator, and
the constant $\Delta$ represents the pairing gap whose value is fixed
by $\Delta = G \langle P^{\dagger}\rangle$ in terms of coupling
constant $G$ and pair moment $P^{\dagger}$ defined by
\begin{equation}
P^{\dagger} = \sum_{\nu >0} a^{\dagger}(\bar{\nu})a^{\dagger}(\nu).
\label{eq_pair}
\end{equation}
Using the Bogoliubov-Valatin transformation \cite{BV},
we can diagonalize the hamiltonian  $H^{\prime} = H_{\rm WS} + V_{\rm 
pair} -
\lambda N_p$, where $H_{\rm WS}$
is the WS hamiltonian, $N_p$ the number operator for the protons, and
$\lambda$  the chemical potential.
Parameters $\Delta$ and $\lambda$ are fixed by the self-consistency 
between
$\Delta$ and $P^{\dagger}$
\begin{mathletters}
\label{eq_BCS}
\begin{equation}
\Delta = G \sum_{\nu>0} u(\nu)v(\nu),
\end{equation}
and
by the constraint on the expectation value of the particle number,
$Z=\langle N_p \rangle$:
\begin{equation}
Z = 2 \sum_{\nu>0} v(\nu)^2.
\end{equation}
\end{mathletters}
In Eq.~(\ref{eq_BCS}), $u$ and $v$ denote the coefficients of the 
Bogoliubov-Valatin transformation
\cite{BV} given by
\begin{mathletters}
  \label{eq_uv}
\begin{equation}
  u(\nu) = 2^{-1/2}\left( 1 +
    \frac{e(\nu)-\lambda}{E(\nu)}\right)^{1/2}, \label{eq_u}
\end{equation}
\begin{equation}
  v(\nu) = 2^{-1/2}\left( 1 -
    \frac{e(\nu)-\lambda}{E(\nu)}\right)^{1/2} \label{eq_v}
\end{equation}
\end{mathletters}
with
\begin{equation}
E(\nu)=\left(\left(e(\nu)-\lambda\right)^2+\Delta^2 \right)^{1/2},
\end{equation}
where $e(\nu)$ stands for the single particle energy of the eigenstate
$\nu$ of the WS hamiltonian.

\begin{figure}

  \epsfxsize=10cm
  \centerline{\epsfbox{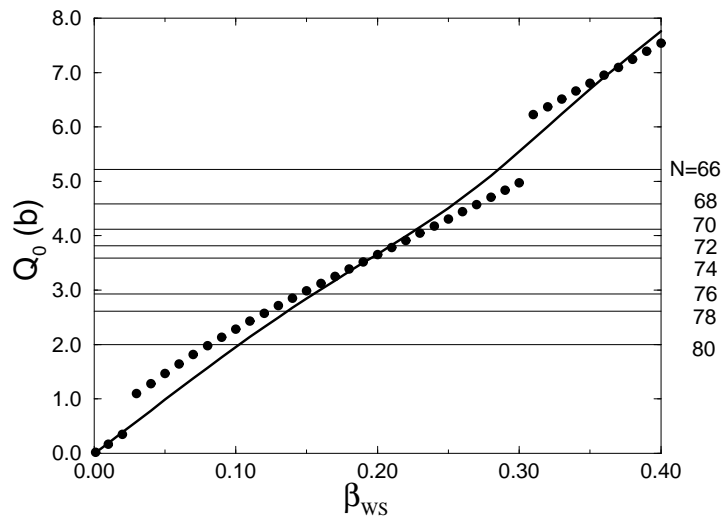}}
  \caption{
    $Q_{\pi 0}$ .vs. $\beta_{\hbox{\scriptsize WS}}$ plot for Ba.
    Proton quadrupole moment $Q_{\pi 0}$ is plotted as a
    function of deformation parameter $\beta_{\hbox{\scriptsize WS}}$ 
of the
    axially-symmetric deformed Woods-Saxon potential.
    The solid curve denotes $Q_{\pi 0}$(WS+BCS) which includes the
    pairing correlation.
    The filled circles express $Q_{\pi 0}$(WS), the quadrupole
    moment calculated with the vanishing pairing gap
    $\Delta_\pi=0$.
    Eight horizontal lines in the figure
    correspond to the experimental data.
    The height of each horizontal line
    is equal to the experimental quadrupole moment for each
    Ba isotope with the neutron number $N$ ranging
    from $N=66$ to 80.
For the sake of clarification,
we have not included the errors of the data. The experimental values for
    $Q_{\pi 0}$ are
    5.2 $\pm$ 0.2,
    4.6 $\pm$ 0.1,
    4.1 $\pm$ 0.1,
    3.8 $\pm$ 0.1,
    3.6 $\pm$ 0.2,
    2.9 $\pm$ 0.1,
    2.62 $\pm$ 0.02,
    2.00 $\pm$ 0.01 (b)
    for 66, 68, $\ldots$, 80, respectively \protect\cite{Ba}.
    }
  \label{fig:Q}
\end{figure}

We determined the $G$ value for ${}^{122}$Ba by the condition that the
pairing gap for the protons should equal the empirical value given by
\cite{BM2}: $\Delta=12/\sqrt{A}$ $\sim$ 1.1 MeV at 
$\beta_{\hbox{\scriptsize WS}}$=0.3, which
$\beta_{\hbox{\scriptsize WS}}$ value roughly corresponds to the 
deformation of Ba with $N=66$
as will be shown below.
The resultant value
is $G=0.16$ MeV.  In the summations with respect to $\nu$ in 
Eq.~(\ref{eq_BCS}),
we adopted all the WS eigenstates with their eigenenergies below the
barrier height \cite{WS}. We note here that the value for $G$ depends
on the number of $\nu$ in these summations.  In this paper, we assume
$G$ being independent of $N$, for simplicity. We have no free
parameter except for $\beta_{\hbox{\scriptsize WS}}$, i.e., the 
quadrupole deformation
parameter of the WS potential.  The value for $\beta_{\hbox{\scriptsize 
WS}}$ will be
determined later by the experimental values of the intrinsic $Q_0$
moment. Refer to Fig.~\ref{fig:Q}.

In terms of intrinsic quadrupole moment $Q_0$ for protons,
we can approximately express B(E2:$0_1^+\rightarrow 2_1^+$) \cite{BM2} 
as
\begin{equation}
B(E2:0_1^+ \rightarrow 2_1^+)=\frac{5}{16\pi}e^2Q_0^2,
\end{equation}
with $e$ being the proton charge.
This is a good approximation for the axially symmetric well deformed
nuclei \cite{BM2}.
The quadrupole moment is given in terms of the $v$ coefficient of 
Eq.~(\ref{eq_v}) by
\begin{equation}
Q_{0} = 2 \sum_{\nu>0} q(\nu) v(\nu)^2, \label{eq_QBCS}
\end{equation}
where $q(\nu)$ represents the quadrupole moment of the WS eigenstate
$\nu$:
\begin{equation}
q(\nu)=\langle \nu | \sqrt{\frac{16\pi}{5}} r^2 Y_{2,0} | \nu \rangle.
\label{eq_q}
\end{equation}
We can thus compare directly the data with the results obtained by the
WS hamiltonian plus the BCS method.

The intrinsic $Q_0$ moment is plotted as a function of 
$\beta_{\hbox{\scriptsize WS}}$ in
Fig.~\ref{fig:Q}. The solid curve represents the quadrupole moment 
calculated
using Eq.~(\ref{eq_QBCS}), which we call $Q$(WS+BCS).
The filled circles denote the quadrupole moment calculated without the
pairing interaction, which we call $Q$(WS).

In general, the quadrupole moment $Q$(WS) increases smoothly with
$\beta_{\hbox{\scriptsize WS}}$ because each occupied orbit of the 
ground state gains the
quadrupole moment gradually with $\beta_{\hbox{\scriptsize WS}}$. It 
also happens that $Q$(WS)
increases suddenly at a certain value of $\beta_{\hbox{\scriptsize 
WS}}$ owing to a change
in the ground state configuration.
In the plot such as the Nilsson diagram \cite{BM2}, where the
level energies are plotted as a function of
the potential deformation, the change in the ground state
configuration can be expected to occur at level crossings of occupied
and unoccupied levels at the current chemical potential.

It is clearly seen from Fig.~\ref{fig:Q} that $Q$(WS) jumps twice at
$\beta_{\hbox{\scriptsize WS}}$ $\sim$0.025 and 
$\beta_{\hbox{\scriptsize WS}}$ $\sim$0.30. These jumps correspond
to the level crossing where the wavefunction of the
ground state changes its configuration.
The former jump is due to the level crossing between the single
particle orbits originating from the spherical sub-shells of $1d_{5/2}$ 
and
$0g_{7/2}$.
This jump is unimportant because the jump in $Q_0$ is too small to 
survive after the pairing
force is switched on as is seen from the curve $Q$(WS+BCS).
The jump at $\beta_{\hbox{\scriptsize WS}}$ $\sim$ 0.3 in $Q$(WS) is 
much bigger than
the former one, and is due to the level crossing between the levels
originating from the unique parity sub-shells,
$0g_{9/2}$ and $0h_{11/2}$, which correspond to the levels with the
Nilsson ``asymptotic'' quantum numbers (Ref.~\cite{BM2})
$\Omega[N_t n_z \Lambda]=\frac{9}{2}^{+}[404]$ and 
$\frac{1}{2}^{-}[550]$,
respectively, where $\Omega$ is the absolute value of the $z$ component 
of the
total angular momentum, $N_t$ the total number of quanta of the
deformed harmonic oscillator, $n_z$ the number of quanta in the 
oscillation
along the symmetry axis of the same oscillator, and $\Lambda$ the 
component of
the orbital angular momentum along the symmetry axis \cite{BM2}.

Hereafter, we denote these two orbits as
$\nu_1=\frac{9}{2}^{+}[404]$ and $\nu_2=\frac{1}{2}^{-}[550]$, 
respectively.
At $\beta_{\hbox{\scriptsize WS}}$ = 0.3, two protons, which occupy the 
$\nu_1$ and $\bar{\nu}_1$ orbits
for $\beta_{\hbox{\scriptsize WS}}$ $<$ 0.3, move to the $\nu_2$ and 
$\bar{\nu}_2$ orbits
which are unoccupied by protons for $\beta_{\hbox{\scriptsize WS}}$ $<$ 
0.3.
By this movement of two protons, $Q$(WS) increases sharply at 
$\beta_{\hbox{\scriptsize WS}}$
= 0.3 by
\begin{equation}
\Delta Q_0 = 2 \left( q(\nu_2) - q(\nu_1) \right) = 1.12 \hbox{(b)}
\end{equation}
with $q(\nu_1)=-0.16$ (b) and $q(\nu_2)=0.40 $(b).
This magnitude of $\Delta Q_0$ amounts to as much as 20\% of the 
experimental
value $Q_0$=5.2 $\pm$ 0.2 (b) for ${}^{122}$Ba \cite{Ba}.
The $\nu_1$ ($\nu_2$) orbit has the largest magnitude of $q$ in 
Eq.~(\ref{eq_q})
with a
negative (positive) sign among the levels with the same sign of $q$ 
near and
below the Fermi surface. Therefore, emptying the $\nu_1$ orbit and 
filling
the $\nu_2$ orbit strongly affect the $Q_0$ value.

We now consider the effect of the pairing force on $Q_0$.  The current
pairing force strength $G$ yields the pairing gap $\Delta=1.34, 1.25,
0.99, 1.09, 1.06$ (MeV) for $\beta_{\hbox{\scriptsize WS}}$=0.0, 0.1, 
0.2, 0.3, 0.4,
respectively.  Due to the level crossing between $\nu_1$ and $\nu_2$,
$\Delta$ increases as $\beta_{\hbox{\scriptsize WS}}$ approaches 0.3.  
Owing to the pairing
interaction, the sharp jumps existing in $Q$(WS) get smoother in 
$Q$(WS+BCS).
The effect of the pairing force on $Q_0$ depends on 
$\beta_{\hbox{\scriptsize WS}}$. The
force reduces $Q_0$ for $\beta_{\hbox{\scriptsize WS}}$ $<$ 0.2 whereas 
it enhances $Q_0$ for
0.2 $<$ $\beta_{\hbox{\scriptsize WS}}$ $<$ 0.3.  The pairwise proton 
transfer takes place
from $\Omega=9/2 (g_{9/2})$ to $\Omega=1/2 (h_{11/2})$, and the
probability of the pairwise transfer increases as 
$\beta_{\hbox{\scriptsize WS}}$ approaches
the level crossing point at $\beta_{\hbox{\scriptsize WS}}$=0.3.  The 
quadrupole moment
$Q$(WS+BCS) thus gets bigger than $Q$(WS) for 0.2 $<$ 
$\beta_{\hbox{\scriptsize WS}}$ $<$ 0.3.

Eight horizontal lines are drawn in Fig.~\ref{fig:Q}, which correspond 
to the
experimental $Q_0$ moments for the Ba isotopes with $N$ from 66 to 80.
Namely, the height of each line is equal to the experimental value of 
the
$Q_0$ moment.
Thus, from the intersection point between each
horizontal line and the solid curve $Q$(WS+BCS), we can determine
$\beta_{\hbox{\scriptsize WS}}$ corresponding to the experimental $Q_0$ 
moment for the individual
isotope. It then turns out that roughly 15\% of B(E2:$0_1^+\rightarrow 
2_1^+$) for Ba with
$N=66$ is due to the pairwise proton transfer
from $\Omega=9/2 (g_{9/2})$ to $\Omega=1/2 (h_{11/2})$.
These results are not sensitive to choices of the WS parameter sets.
When we used the WS parameter sets in \cite{WS} other than the current
``universal'' set, we got the values ranging from 15\% to 20\%.

The Ba isotopes can be classified into three groups by a relation
between $Q$(WS+BCS) and $Q$(WS):
\begin{center}
\begin{tabular} {llcl}
(a) &  $ Q\mbox{(WS+BCS)} < Q\mbox{(WS)}     $ & for & $N$= 76 $\sim$ 
80 \\
(b) &  $ Q\mbox{(WS+BCS)} \sim Q\mbox{(WS)}  $ & for & $N$= 70 $\sim$ 
74 \\
(c) &  $ Q\mbox{(WS+BCS)} > Q\mbox{(WS)}     $ & for & $N$= 66 $\sim$ 
68 \\
\end{tabular}
\end{center}
\noindent We should here note that
$\Delta$ for (b) is as big as 1 MeV although $Q$(WS+BCS) $\sim$ $Q$(WS).

A comparison between Figs.~\ref{fig:Ba} and \ref{fig:Q} leads us to
a correspondence between the results of IBM-2 and of the WS hamiltonian 
plus
the BCS method:
\begin{itemize}
\item[(a)]
  IBM-2 with the Pauli factor gives B(E2) closer to the
  experimental data than the IBM-2 without the Pauli factor.
\item[(b)]
  IBM-2 without the Pauli factor gives B(E2) closer to the
  experimental data than the IBM-2 with the Pauli factor.
\item[(c)] Both IBM-2 with and without the Pauli factor
underestimate B(E2).
\end{itemize}

The OAI mapping used in the derivation of the IBM-2 E2 operator in
Eqs.~(\ref{eq_e2}-\ref{eq_chi}) with the Pauli factor is based on the
seniority scheme \cite{OAI}. This scheme is suitable for the case in
which the single particle levels are nearly degenerate regarding
energy at zero deformation.  The OAI mapping thus does not take
account of the intruder orbit $\Omega=9/2^+(g_{9/2})$ which intrudes the
$Z=50-82$ shell originating at zero deformation from below the energy
gap at $Z=50$.  It is thus reasonable that the IBM-2 predictions
underestimate B(E2:$0_1^+\rightarrow 2_1^+$) at $N=66, 68$ while the 
IBM-2 predictions with the
Pauli factor are in good agreement with the experimental data for the
less deformed Ba isotopes with $N\ge 76$.

We should mention here the Xe ($Z=54$) and Ce ($Z=58$) isotopes,
though the results for Xe and Ce are not shown in detail in this
paper.  We have applied the WS hamiltonian plus the BCS method to the 
Xe and
Ce isotopes in a way similar to the procedures used for Ba.  It then
turns out that, if we chose the Chepurnov set (the ``universal'' set)
in \cite{WS}, the contribution of the level crossing between
the orbits originating from $\Omega=9/2(g_{9/2})$ and the 50-82 shell
to B(E2:$0_1^+\rightarrow 2_1^+$) for Xe with $N=66$ is roughly 20\% 
(10\%).
The corresponding quantity for Ba with $N=66$ is roughly 20\% (15\%).
For Ce with $N=66$, the effect of the level crossings on 
B(E2:$0_1^+\rightarrow 2_1^+$) is
smaller than 5\% when we ignore the error in the experimental data.

The B(E2:$0_1^+\rightarrow 2_1^+$) value predicted by the Grodzins rule 
in Eq.~(\ref{eq_grodzins})
is only 75\% of the experimental data for Xe with $N=66$,
while the prediction for Ce with $N=66$ lies
within the error bar of the experimental B(E2:$0_1^+\rightarrow 2_1^+$) 
data \cite{Ce}.

The B(E2:$0_1^+\rightarrow 2_1^+$) values calculated with IBM-2 without 
the Pauli factor are
nearly the same as the experimental data for Xe with $N=64-68$
\cite{raman95} and Ce with $N=66-74$, namely for the isotopes with
relatively large B(E2) values.  On the other hand, for the less
deformed nuclei such as Xe with $N=70-80$ and Ce with $N=76$, IBM-2
with the Pauli factor gives the B(E2:$0_1^+\rightarrow 2_1^+$) values 
closer to the data
than IBM-2 without the Pauli factor. These trends are thus consistent
with the IBM-2 results for the Ba isotopes in Fig.~\ref{fig:Ba}.  This 
fact
suggests that some unknown mechanism cancels the effect of the Pauli
factor in Eqs.~(\ref{eq_Q}-\ref{eq_chi}), and this plays a role
mainly for the well deformed nuclei.  To make this mechanism clear is
beyond the scope of the current paper.  We should investigate this
problem further in the future.

\section{Summary}
By the use of the Woods-Saxon hamiltonian plus the BCS method, we 
calculated the
intrinsic $Q_0$ moment as a function of the deformation parameter of
the WS potential.
Comparing the calculated $Q_0$ moment with the
experimental data, we can determine the value of 
$\beta_{\hbox{\scriptsize WS}}$
for each Ba isotope.  It leads us to find a detailed way how protons 
occupy the
deformed single particle orbits.
By this procedure, we found
that as much as 15\% of B(E2:$0_1^+\rightarrow 2_1^+$) for ${}^{122}$Ba 
is due to the
pairwise transfer from the proton intruder orbit originating from
$\Omega=9/2^+ (g_{9/2})$ to the proton unique-parity orbit originating
from $\Omega=1/2^- (h_{11/2})$.
We have shown that the contribution to $Q_0$
from this pairwise transfer
is roughly equal to the
deviation of the experimental data from the theoretical predictions by
IBM-2 without the Pauli factor. Similar deviations are found for the
predictions of the Grodzins rule.
We also found that this contribution to B(E2:$0_1^+\rightarrow 2_1^+$) 
decreases rapidly as
$N$ goes away from the mid-shell value of $N=66$.  This fact suggests
that the pairwise transfer from the orbit with $\Omega=9/2^+ (\pi
g_{9/2})$ to the one with $\Omega=1/2^- (\pi h_{11/2})$ is responsible
for the sharp increase of B(E2:$0_1^+\rightarrow 2_1^+$) observed in 
the Ba isotope with
$N=66$.

One of the authors (K.U.) would like to acknowledge the Junior Research
Associate Program of Japan Science and Technology Agency.

\end{document}